\documentstyle[amssymb,twocolumn,prl,aps]{revtex}
\newcommand{\ket}[1]{{|#1\rangle}}
\newcommand{\bra}[1]{{\langle#1|}}

\begin{document}

\title{A Nonadditive Quantum Code}

\author{Eric M. Rains, R. H. Hardin, Peter W. Shor, and N. J. A. Sloane}
 
\address{AT\&T Research, 600 Mountain Ave., 
Murray Hill, NJ 07974, USA}

\date{March 3, 1997}
\maketitle


\begin{abstract}

Up to now every good quantum error-correcting code discovered has had
the structure of an eigenspace of an Abelian group generated by
tensor products of Pauli matrices; such codes are known as
{\em stabilizer} or {\em additive} codes.  In this letter we 
present the first example of a code that is better than any
code of this type.  It encodes six states in five qubits and
can correct the erasure of any single qubit.

\end{abstract}


\narrowtext



\renewcommand{\arraystretch}{.33}

Most present models for quantum computers require quantum error-correcting
codes \cite{Shor,Steane,EM,KL} for their operations.  Many recent papers have been
devoted to the construction of these codes \cite{Steane2,CS,Gott,crss,crss2}.  
Up to now all good codes known have
fallen into the class of what have been called ``stabilizer'' or ``additive''
codes \cite{Gott,crss2}.  These are not the most general quantum codes, 
however, and
it was an open question as to whether more general codes could be more
powerful.  In the present letter we present a code which is not a
stabilizer code, and which is strictly better than any
code of this type. 

Stabilizer codes are defined as follows.  Let the extraspecial group 
$E$ be generated by all tensor products
$M_1 \otimes M_2 \otimes \ldots \otimes M_n,$ where each $M_i$ is
one of $\{I, \sigma_x, 
\sigma_y, \sigma_z\}$, where 
\[
I = \left(\begin{array}{cc}  \scriptstyle 1& \scriptstyle 0
\\ \scriptstyle 0 & \scriptstyle 1\end{array}\right), 
\sigma_x = \left(\begin{array}{cc}  \scriptstyle 0& \scriptstyle 1
\\ \scriptstyle 1 & \scriptstyle 0\end{array}\right), \nonumber 
\sigma_y = \left(\begin{array}{cc}  \scriptstyle 0& \scriptstyle -i
\\ \scriptstyle i & \scriptstyle 0\end{array}\right), 
\sigma_z = \left(\begin{array}{cc}  \scriptstyle 1& \scriptstyle 0
\\ \scriptstyle 0 & \scriptstyle -1\end{array}\right), 
\]
$\sigma_x$, $\sigma_y$, $\sigma_z$ being
the Pauli matrices.  Then a {\em stabilizer}
code is a joint eigenspace of an Abelian subgroup of $E$.  
\renewcommand{\arraystretch}{1}

\paragraph*{The new code.} 
Since a quantum code is a subspace of the Hilbert space ${\Bbb C}^{2^n}\!$,
we may define it by giving the orthogonal projection matrix onto this
subspace.  Since the Pauli matrices together with $I$ form a basis
for the space of $2\times 2$ Hermitian matrices, we may write the
projection as a linear combination of elements of $E$.  In this form, the
new code $Q$ is defined by the projection matrix
\begin{eqnarray}
P = \scriptsize{1}/\scriptsize{16} & \big[ & 3\, \phantom{(}I \otimes I \otimes I \otimes I \otimes I \nonumber\\
&+& \phantom{2\,} (I \otimes \sigma_z \otimes \sigma_y \otimes \sigma_y \otimes \sigma_z
)_{\rm cyc} \nonumber\\
&+& \phantom{2\,} (I \otimes \sigma_x \otimes \sigma_z \otimes \sigma_z \otimes \sigma_x
)_{\rm cyc} \nonumber\\
&-& \phantom{2\,} (I \otimes \sigma_y \otimes \sigma_x \otimes \sigma_x \otimes \sigma_y
)_{\rm cyc} \nonumber\\
&+& 2\, (\sigma_z \otimes \sigma_x \otimes \sigma_y \otimes \sigma_y \otimes \sigma_x
)_{\rm cyc} \nonumber\\
&-& 2 \phantom{(\,}\sigma_z \otimes \sigma_z \otimes \sigma_z \otimes \sigma_z \otimes \sigma_z
\big], \nonumber
\end{eqnarray}
where the subscript ``cyc'' indicates that all five cyclic shifts
occur.  It is straightforward to verify that $P^2 = P$ and 
${\rm Trace}(P) = 6$, so that the eigenvalues of $P$ are 1 (6 times) and
0 (26 times).  Thus P projects onto a six-dimensional space.

It remains to verify that the code has minimum distance 2, that is, that
it can correct one erasure.  For a code to correct one erasure it
suffices for it to be orthogonal to its image under any single
qubit error.  Since $Q$ is cyclic, it is enough to consider an 
error in the first qubit.  
There are thus three cases to check:  conjugation of $P$ by
$\eta = M \otimes I \otimes I \otimes I \otimes I \otimes I$
where $M$ is one of $\sigma_x$, $\sigma_y$, $\sigma_z$,
obtaining $P' = \eta P \eta$.  We verify in each case that $P P' = 0$.
Thus $Q$ is a ((5,6,2)) code, as claimed.

\paragraph*{Further properties of the code.}
Since the code is cyclic, we label the qubits by
$0,1,\ldots,4$ $({\rm mod\ } 5)$.  The code also has the 
symmetry defined by 
\begin{equation}
k \rightarrow 2k\ ({\rm mod\ } 5),\
\sigma_x \rightarrow \sigma_y,\
\sigma_y \rightarrow -\sigma_x.
\end{equation}
There are further symmetries that do not permute the qubits, for example
conjugation by the element
\begin{equation}
\sigma_z \otimes \sigma_x \otimes \sigma_y \otimes \sigma_y \otimes \sigma_x
\end{equation}
of $E$, or any of its cyclic shifts.
These five elements generate a subgroup of $E$ of size 32, which we 
call $H$.  Moreover,
since this group is Abelian, it has 32 one-dimensional eigenspaces, and 
$Q$ is the span of six of them.  (Each of these eigenspaces is a
((5,1,3)) stabilizer quantum code.)  This gives an explicit
basis for our code, namely:
\begin{equation}
\ket{00000} -(\ket{00011})_{\rm cyc} 
+(\ket{00101})_{\rm cyc} -(\ket{01111})_{\rm cyc} 
\end{equation}
together with all five cyclic shifts of
\begin{eqnarray}
&&\ket{00001}-\ket{00010}-\ket{00100}-\ket{01000}-\ket{10000}\nonumber\\
&+&\ket{00111}-\ket{01110}-\ket{11100}+\ket{11001}+\ket{10011}\nonumber\\
&-&\ket{01011}+\ket{10110}-\ket{01101}+\ket{11010}-\ket{10101}\nonumber\\
&-&\ket{11111}
\end{eqnarray}
By examining how single bit errors act on these 32 eigenspaces, one
can again see that the image of $Q$ under such an error is orthogonal
to $Q$, by noting that the six new eigenspaces are disjoint from the
original six.

The symmetries discussed above generate a group of order 640.  There 
is an additional symmetry, which can be described as follows: 
First, permute the columns as $k\rightarrow k^3$, that is, exchange the
qubits 2 and 3.  Next, for each qubit, negate one of the Pauli matrices
and exchange the other two, where the Pauli matrices negated are 
$\sigma_z$, $\sigma_y$, $\sigma_x$, $\sigma_x$, $\sigma_y$, respectively.  
This increases the size of the symmetry group to 3840.
The group acts as the permutation group $S_5$ on the qubits.  This is 
the full group of symmetries of the code, that is, the full subgroup of 
the semidirect product of $S_5$ and $PSU(2)^5$ that preserves the 
code \cite{Rains2}.

In principal one can construct other codes in a similar manner, e.g.
as the spans of translates of self-dual stabilizer codes.
Let $C_0$ be a self-dual additive code of length $n$ with associated 
stabilizer quantum code $Q_0$ and let $C$ be the union of $K$ cosets 
of $C_0$.  If $C$ has minimum distance $d$, then there exists 
an $((n,K,d))$ quantum code.  This code is obtained as the span of the 
translates of $Q_0$ under the operators in $C$.  For our code, for
instance, $C_0$ is the $[[5,0,3]]$ additive code $H$ and $C$ is
$C_0 + (\sigma C_0)_{\rm cyc}$, where 
\[
\sigma = \sigma_x \otimes \sigma_z \otimes I \otimes I \otimes I.
\]
Thus, for example, we have
\begin{equation}
P = P_0 + (\sigma P_0\sigma^{-1})_{\rm cyc},
\end{equation}
where $P_0$ is $\ket{x}\bra{x}$ and $\ket{x}$ is given by Eq.~(3).

We remark that the only stabilizer code that contains $Q$ is the trivial
code consisting of the entire Hilbert space.

\paragraph*{How the code was discovered.}
Investigation of the linear programming bound for quantum codes
(cf. \cite{crss,ShoLaf,Rains}) revealed that for a length 5 code
of minimum distance 2 the dimension $K=6$ is extremal.  In fact,
the inequalities cannot even be satisfied for any {\em real} number
$K > 6$.  This suggested a ((5,6,2)) code as a natural place to look.
Further, the weight enumerator of any such code is uniquely determined
by linear programming; with the normalization used in \cite{Rains}, 
it is 
\begin{equation}
A(u,v) = 36u^5 + 60 uv^4 + 96v^5.
\end{equation}
The best stabilizer code of this length and distance has dimension $K=4$ 
\cite[Table III]{crss}.

Rather than look for a projection matrix with this weight enumerator,
we decided to look for a matrix with this weight enumerator that 
happened to be a projection.  
We describe a near projection matrix $M$ as a linear combination
of elements of $E$.  
Beginning with a random matrix $M$, we first scaled the contribution 
from the terms of each weight so as to force the desired weight 
enumerator.  We then replaced $M$ by $M' = 2M^2-M^4$.  (This 
moves small eigenvalues of M closer to 0 and large eigenvalues
closer to 1.)  We then iterated these two steps.  On our first attempt, 
the algorithm converged to a (complex) projection matrix.  
Since the algorithm preserves symmetry, by choosing our initial
guess to be real, we obtained a real code
equivalent to the code described above.  We omit the details of 
transforming the computer output into the above form.
Note that we did not choose our initial guess to be cyclic, but 
nonetheless obtained a code equivalent to one with cyclic symmetry.

\paragraph*{Acknowledgment.}
We thank Rob Calderbank for helpful discussions concerning the
description of the code in terms of cosets.

\end{document}